\newcommand{\D}{\mbox{\bf D}}
\newcommand{\Z}{\mbox{\bf Z}}

\newcommand{\cW}{{\cal W}}

\newcommand{\iso}{\mathop{\mathrm{Iso}}}
\newcommand{\sgn}{\mathop{\mathrm{sgn}}}

\documentstyle[12pt,psfig]{elsart}

\renewcommand{\Rset}{\mbox{\bf R}}

\begin{document}
\begin{frontmatter}

\title{Cycling chaos: its creation, persistence and loss of stability in a
model of nonlinear magnetoconvection}

\author{Peter~Ashwin}
\address{Department of Mathematical and Computing Sciences, University of
Surrey, Guildford GU2 5XH, UK}
\and
\author{A.M.~Rucklidge}
\address{Department of Applied Mathematics and Theoretical Physics,
University of Cambridge, Silver Street, Cambridge CB3 9EW, UK}

\begin{abstract}
 We examine a model system where attractors may consist of a heteroclinic cycle
between chaotic sets; this `cycling chaos' manifests itself as trajectories
that spend increasingly long periods lingering near chaotic invariant sets
interspersed with short transitions between neighbourhoods of these sets. This
behaviour can be robust (i.e., structurally stable) for systems with symmetries
and provides robust examples of non-ergodic attractors in such systems; we
examine bifurcations of this state.

We discuss a scenario where an attracting cycling chaotic state is created at a
blowout bifurcation of a chaotic attractor in an invariant subspace. This is a
novel scenario for the blowout bifurcation and causes us to introduce the idea
of {\em set supercriticality\/} to recognise such bifurcations. The robust
cycling chaotic state can be followed to a point where it loses stability at a
resonance bifurcation and creates a series of large period attractors.

The model we consider is a $9$th order truncated ordinary differential equation
(ODE) model of three-dimensional incompressible convection in a plane layer of
conducting fluid subjected to a vertical magnetic field and a vertical
temperature gradient.  Symmetries of the model lead to the existence of
invariant subspaces for the dynamics; in particular there are invariant
subspaces that correspond to regimes of two-dimensional flows. Stable
two-dimensional chaotic flow can go unstable to three-dimensional flow via the
cross-roll instability. We show how the bifurcations mentioned above can be
located by examination of various transverse Liapunov exponents. We also
consider a reduction of the ODE to a map and demonstrate that the same
behaviour can be found in the corresponding map. This allows us to describe and
predict a number of observed transitions in these models.
 \end{abstract}


\end{frontmatter}

\section{Introduction}

There has been a lot of recent interest in the chaotic dynamics of nonlinear
systems that possess invariant subspaces; a number of quite subtle dynamical
effects come to light in examining the interaction of attractors with the
invariant subspaces. Moreover a good understanding of these effects is
essential in interpreting and predicting dynamics of simulations and
experiments where the presence of discrete spatial symmetries implies the
existence of invariant subspaces.

A fundamental bifurcation in such a setting is the blowout bifurcation
\cite{Ott&Som94} where a chaotic attractor within  an invariant subspace loses
stability to perturbations transverse to the invariant subspace. In doing so,
can either create a nearby `branch' of chaotic attractors or lose stability
altogether.

In contrast with this, the presence of invariant subspaces can lead to the
existence of what have been called robust heteroclinic cycles \cite{Guc&Hol88}
between equilibria, that is, heteroclinic cycles that are persistent under
small perturbations that preserve the symmetry. These cycles may or may not be
attracting \cite{KrupaMelbourne}. Recently it has been recognised that cycles
to more complicated invariant sets can also occur robustly in symmetric
systems, in particular to chaotic invariant sets; this behaviour was named
`cycling chaos' in a recent paper by Dellnitz {\em et al.}\ \cite{Del&al95} and
has been further investigated by Field \cite{Fie96} and Ashwin \cite{Ash97}.

In this paper we find there is a connection between these dynamical properties;
we show a scenario where a blowout bifurcation creates an attracting `cycling
chaotic' state in a bifurcation that is analogous to a saddle-node homoclinic
bifurcation with equilibria replaced by chaotic invariant sets. We also
investigate how the attracting cycling chaotic state that is created in the
blowout bifurcation loses stability at a resonance of Liapunov exponents. (A
{\em resonance bifurcation\/} in its simplest form occurs when a homoclinic
cycle to an equilibrium loses attractiveness; this occurs when the real parts
of eigenvalues of the linearisation become equal in magnitude \cite{Cho&al90}.)
In spite of the system being neither a skew product nor being a homoclinic
cycle to a chaotic set as in \cite{Ash97} we see similar behaviour and can
predict the loss of stability by looking at a rational combination of Liapunov
exponents.

We find this scenario of a blowout bifurcation to cycling chaos is a mechanism
for transition from stable two-dimensional to fully three-dimensional
magnetoconvection. The model we study is a Galerkin truncation for
magnetoconvection in a region with square plan, subject to periodic boundary
conditions on vertical boundaries and idealised boundary conditions on
horizontal boundaries. Phenomenologically speaking we see a change from a
chaotically varying two-dimensional flow (with trivial dependence on the third
coordinate, and which comes arbitrarily close to a trivial conduction state) to
an attracting state where trajectories spend increasingly long times near one
of two symmetrically related two-dimensional flows interspersed with short
transients. We explain and investigate this transition in terms of a blowout
bifurcation of a chaotic attractor in an invariant subspace.

In the paper of Ott and Sommerer \cite{Ott&Som94} that coined the phrase
`blowout bifurcation', two scenarios are identified. Either the blowout was
{\em supercritical\/} in which case it leads to an {\em on-off intermittent\/}
state \cite{Pla&al93}, or it is {\em subcritical\/} and there is no nearby
attractor after the bifurcation. We find an additional robust possibility for
bifurcation at blowout.

Near this transition the three-dimensional flow patterns show characteristics
of intermittent cycling between two symmetrically related `laminar' states
corresponding to two-dimensional flows, but the time spent near the laminar
state is, on average, infinite. This suggests that the blowout is
supercritical, but in a weaker sense which we make precise.  Namely, we say a
blowout is {\em set supercritical\/} if there is a branch of chaotic attractors
after the blowout whose limit contains the attractor in the invariant subspaces
before the blowout. In particular there may be other invariant sets contained
in this limit and so any natural measures on the bifurcating branch of
attractors (if they exist) need not limit to the natural measure of the system
on the invariant subspace.

We also show that the attractors corresponding to two-dimensional flows are not
Liapunov stable, but are Milnor attractors near the transition to three
dimensions, and so in particular we expect the presence of noise to destabilise
two-dimensional attractors near blowout by a {\em bubbling\/} type of mechanism
\cite{AshBueSte94}.

We find in our example that the state of cycling chaos is attracting once it
has been created: trajectories cycle between neighbourhoods of the chaotic sets
within the invariant subspaces, and the time between switches from one
neighbourhood to the next increases geometrically as trajectories get closer
and closer to the invariant subspaces. By estimating the rate of increase of
switching times, we are able to show that cycling chaos ceases to be attracting
in a resonance bifurcation. One remarkable aspect of this study is that we are
able to predict the parameter values at which the blowout bifurcation and the
resonance occur, requiring only a single numerical average over the chaotic set
within the invariant subspace.

The paper is organised as follows: in section~\ref{secmodel} we introduce the
ODE model for magnetoconvection, discuss its symmetries and corresponding
invariant subspaces. This is followed by a description of the creation,
persistence and loss of stability of the cycling chaos on varying a parameter
in numerical simulations in Section~\ref{secnumerics}. Section~\ref{secmap}
shows how one can, under certain assumptions, derive a map model of the
dynamics of the ODE that has the same dynamical behaviour.
Section~\ref{secblowout} is a theoretical analysis of the blowout bifurcation
that creates the cycling chaotic attractor and is followed in
Section~\ref{secresonance} by a theoretical analysis of its loss of stability.
Finally, Section~\ref{secdiscuss} discusses some of the implications of this
work on the chaotic dynamics of symmetric systems.

\section{An ODE model for magnetoconvection}
\label{secmodel}

The model we study is an ODE on $\Rset^9$ described by the following equations
 \begin{eqnarray}\label{eqode}
 \dot{x}_0&=&    \mu x_0 + x_0\theta - x_2 x_1 - \beta y_0^2 x_0, \nonumber\\
 \dot{x}_1&=&   -\nu x_1 + x_0 x_2,                               \nonumber\\
 \dot{x}_2&=&   -\sigma x_2 - \sigma Q a + \gamma x_0 x_1,        \nonumber\\
 \dot{y}_0&=&    \mu y_0 + y_0\theta - y_2 y_1 - \beta x_0^2 y_0, \nonumber\\
 \dot{y}_1&=&   -\nu y_1 + y_0 y_2,                                        \\
 \dot{y}_2&=&   -\sigma y_2 - \sigma Q b + \gamma y_0 y_1,        \nonumber\\
 \dot{a}&=&      \zeta(x_2 - a),                                  \nonumber\\
 \dot{b}&=&      \zeta(y_2 - b),                                  \nonumber\\
 \dot{\theta}&=&-\theta - x_0^2 - y_0^2.                          \nonumber
 \end{eqnarray}
These ODEs have been derived as an asymptotic limit of a model of
three-dimensional incompressible convection in a plane layer, with an imposed
vertical magnetic field; for further details and details of its derivation, see
\cite{Metal96,RM95,RM96}. In the context of this model, $x_0$ and $y_0$
represent the amplitudes of convective rolls with their axes aligned in the $y$
and $x$ (horizontal) directions respectively, $x_1$ and $y_1$ represent modes
that cause the rolls to tilt, and $x_2$ and $y_2$ represent shear across the
layer in the $x$ and $y$ directions. The modes $a$ and $b$ represent the
horizontal component of the magnetic field in the $x$ and $y$ directions, and
$\theta$ represents the horizontally averaged temperature.

The model has are five primary parameters:  $\mu$~is proportional to the
imposed temperature difference across the layer, with $\mu=0$ at the initial
bifurcation to convection; $\beta$~is related to the horizontal spatial
periodicity length, but is an arbitrary small parameter in the model of
\cite{Metal96,RM95}; $\sigma$~and $\zeta$ are dimensionless viscous and
magnetic diffusion coefficients; and $Q$~is proportional to the square of the
imposed magnetic field. Note that $\sigma$, $\zeta$ and $Q$ are scaled by
factors of $4$, $4$ and $\pi^2$ from their usage in \cite{Metal96,RM95,RM96}.
Two secondary parameters that we use are $\gamma=3(1+4\sigma)/16\sigma$ and
$\nu=(9\sigma/(1+4\sigma))-\mu$. In the parameter regime of interest, all
parameters are non-negative.

\subsection{Symmetries of the model}

Consider $\D_4\dot{+}T^2$ acting on the plane with unit cell $[0,2\pi)^2$ in
the usual way, with the torus $T^2$ acting by translations on the plane, and
$\D_4$ by reflections in the axes and rotation through $\pi/2$. We define the
following group elements
 \begin{equation}
 \begin{tabular}{rll}
 $\kappa_x\colon{}$  & reflection through $x=0$          &
                              $(x,y)\mapsto(-x,y)$,                  \\
 $\kappa_x'\colon{}$ & reflection through $x=\pi/2$      &
                              $(x,y)\mapsto(\pi-x,y)$,               \\
 $\kappa_y\colon{}$  & reflection through $y=0$          &
                              $(x,y)\mapsto(x,-y)$,                  \\
 $\kappa_y'\colon{}$ & reflection through $y=\pi/2$      &
                              $(x,y)\mapsto(x,\pi-y)$,               \\
 $\rho\colon{}$      & rotation about $(x,y)=(\pi,\pi)$ \qquad &
                              $(x,y)\mapsto(2\pi-y,x)$,              \\
 $\tau_{(\xi,\eta)}\colon{}$ & translation               &
                              $(x,y)\mapsto(x+\xi,y+\eta)$. 
 \end{tabular}
 \end{equation}
Note that $\rho$, $\tau_{(\xi,\eta)}$ and any reflection $\kappa$ can be used 
to generate the group $\D_4\dot{+}T^2$. 

 \begin{table}
 \caption{Selected fixed point subspaces $S$ of the action of $G$ on $\Rset^9$
together with name, representative point and dimension
of $S$. There are others (e.g. $(0,x,0,0,0,0,0,0,t)$) but these are not
important for the dynamics we discuss here.}
 \label{tabiso}
 \begin{tabular}{lllc}
 $S$      & Name          & Representative point   & $\dim S$ \\
 \hline   
 $F$      & Full symmetry & $(0,0,0,0,0,0,0,0,t)$  & 1 \\
 $R_x$    & $x$-rolls     & $(x,0,0,0,0,0,0,0,t)$  & 2 \\
 $R_y$    & $y$-rolls     & $(0,0,0,x,0,0,0,0,t)$  & 2 \\
 $D_+$    & $+$ diagonal  & $(x,0,0,x,0,0,0,0,t)$  & 2 \\
 $D_-$    & $-$ diagonal  & $(x,0,0,-x,0,0,0,0,t)$ & 2 \\
 $R_{xy}$ & Mixed modes   & $(x,0,0,y,0,0,0,0,t)$  & 3 \\
 $P_x$    & $x$-rolls + shear
                          & $(x,y,z,0,0,0,a,0,t)$  & 5 \\
 $P_y$    & $y$-rolls + shear
                          & $(0,0,0,x,y,z,0,a,t)$  & 5 \\
 $Q_x$    & $x$-rolls + shear + crossrolls
                          & $(x,y,z,w,0,0,a,0,t)$  & 6 \\
 $Q_y$    & $y$-rolls + shear + crossrolls
                          & $(w,0,0,x,y,z,0,a,t)$  & 6 \\
 $T$      & No symmetry   & $(u,v,w,x,y,z,a,b,t)$  & 9 \\
 \hline   
 \end{tabular}
 \end{table}

 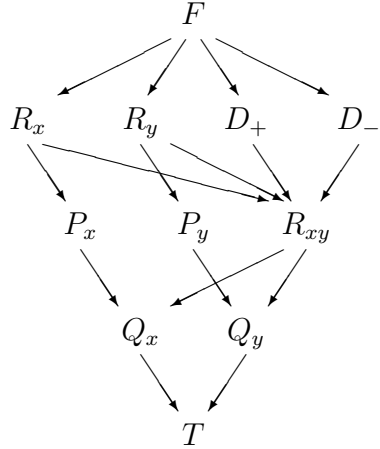
\begin{figure}
 \begin{center}\setlength\unitlength{1mm}\begin{picture}(50,60)
 \put(25,0){\makebox(0,0)[bc]{$T$}}
 \put(18,14){\makebox(0,0)[bc]{$Q_x$}}
 \put(32,14){\makebox(0,0)[bc]{$Q_y$}}
 \put(10,28){\makebox(0,0)[bc]{$P_x$}}
 \put(25,28){\makebox(0,0)[bc]{$P_y$}}
 \put(40,28){\makebox(0,0)[bc]{$R_{xy}$}}
 \put(3,42){\makebox(0,0)[bc]{$R_x$}}
 \put(18,42){\makebox(0,0)[bc]{$R_y$}}
 \put(32,42){\makebox(0,0)[bc]{$D_+$}}
 \put(47,42){\makebox(0,0)[bc]{$D_-$}}
 \put(25,56){\makebox(0,0)[bc]{$F$}}
 \put(18,12){\vector(2,-3){5}}
 \put(32,12){\vector(-2,-3){5}}
 \put(10,26){\vector(2,-3){5}}
 \put(25,26){\vector(2,-3){5}}
 \put(40,26){\vector(-2,-3){5}}
 \put(37,26){\vector(-2,-1){15}}
 \put(3,40){\vector(2,-3){5}}
 \put(18,40){\vector(2,-3){5}}
 \put(33,40){\vector(2,-3){5}}
 \put(47,40){\vector(-2,-3){5}}
 \put(5,40){\vector(4,-1){30}}
 \put(22,40){\vector(2,-1){15}}
 \put(22,54){\vector(-2,-1){15}}
 \put(24,54){\vector(-2,-3){5}}
 \put(26,54){\vector(2,-3){5}}
 \put(28,54){\vector(2,-1){15}}
 \end{picture}\end{center}
 \caption{A portion of the isotropy lattice for the action of $G$ 
 on $\Rset^9$ under which ({\protect\ref{eqode}}) is equivariant. We
 have shown fixed point subspaces of some conjugate subgroups separately for
 clarity. The isotropies of $P_x$ and $P_y$ are the smallest isotropies
 that physically involve only two-dimensional effects.}
 \label{figlattice}
 \end{figure}

We consider the subgroup
 \begin{equation}
 G=\langle\kappa_x,\kappa_x',\kappa_y,\kappa_y',\rho\rangle.
 \end{equation}
Since $G$ contains the subgroup $(\Z_2)^2$, generated by $\kappa_x\kappa_x'$
and $\kappa_y\kappa_y'$, of translations $T^2$, it follows that $G$~is
isomorphic to a semidirect product $\D_4\dot{+}(\Z_2)^2$ ($|G|=32$). The ODE
(\ref{eqode}) is equivariant under the group $G$ of symmetries acting on
$\Rset^9$ by
 \begin{eqnarray}
 \kappa_x (x_0,x_1,x_2,y_0,y_1,y_2,a,b,\theta) 
      = & (x_0,-x_1,-x_2,y_0,y_1,y_2,-a,b,\theta),   \nonumber\\
 \kappa_x' (x_0,x_1,x_2,y_0,y_1,y_2,a,b,\theta) 
       = & (-x_0,x_1,-x_2,y_0,y_1,y_2,-a,b,\theta),           \\
 \rho (x_0,x_1,x_2,y_0,y_1,y_2,a,b,\theta) 
  = & (y_0,-y_1,-y_2,x_0,x_1,x_2,-b,a,\theta).      \nonumber
 \end{eqnarray}
This action gives rise to a number of isotropy types, shown in
Table~\ref{tabiso}. Figure~\ref{figlattice} gives a partial isotropy lattice
for this group action. Dynamics in $F$ always decays to the trivial equilibrium
point, corresponding to the absence of convection. We refer to dynamics in
$P_x$ and $P_y$ as two-dimensional, since these correspond to two-dimensional
convection in the original problem (though $\dim P_x$ is~5). Dynamics in $R_x$
and $R_y$ corresponds to mirror symmetric two-dimensional rolls with their axes
aligned along the $y$ and $x$ directions; we refer to equilibrium points in
these subspaces as $x$-rolls and $y$-rolls respectively. In $P_x$ and $P_y$,
convection is two-dimensional but not mirror symmetric, and is referred to as
tilted rolls. Dynamics in $Q_x$ and $Q_y$ corresponds to three-dimensional
convection that is still invariant under one mirror reflection (tilted rolls
with a cross-roll component). Otherwise we say the dynamics is fully three
dimensional.

In a slight break from convention we say the fixed point subspaces $S$ as
having isotropy subgroup $\iso(S)$ rather than considering the isotropy
subgroups as the fundamental objects.

\section{Numerical simulations of the ODEs}
\label{secnumerics}

We present numerical simulations of the ODEs that demonstrate two aspects of
cycling chaos that we seek to explain: how cycling chaos can be created in a
blowout bifurcation, and how cycling chaos can cease to be attracting.

We concentrate on parameter values that are known to have Lorenz-like chaotic
dynamics within $P_x$ and $P_y$:
 \begin{equation}
\mu=0.1655, ~~Q=1/\pi^2, ~~\sigma=0.125 ~\mbox{ and }~\zeta=0.05
 \end{equation}
(and hence $\nu=0.5845$ and $\gamma=2.25$), although we note that qualitatively
similar attractors are found for a large proportion of nearby parameter values.
These parameter values correspond to those in Figures~15(c) and 20(a) in
\cite{RM96}. The numerical method used was a Bulirsch--Stoer adaptive
integrator \cite{NumRecipes}, with a tolerance for the relative error set to
$10^{-12}$ for each step.

We use the parameter $\beta$ as a normal parameter (see \cite{AshBueSte96}) for
the dynamics in $P_x$ and $P_y$; that is, it controls instabilities transverse
to $P_x$ and $P_y$ in the directions $Q_x$ and $Q_y$ without altering the
dynamics inside $P_x$ or~$P_y$.

\subsection{Cycling chaos}
\label{secnumericscyclingchaos}

 \begin{figure}
 \begin{center}
 \mbox{\psfig{file=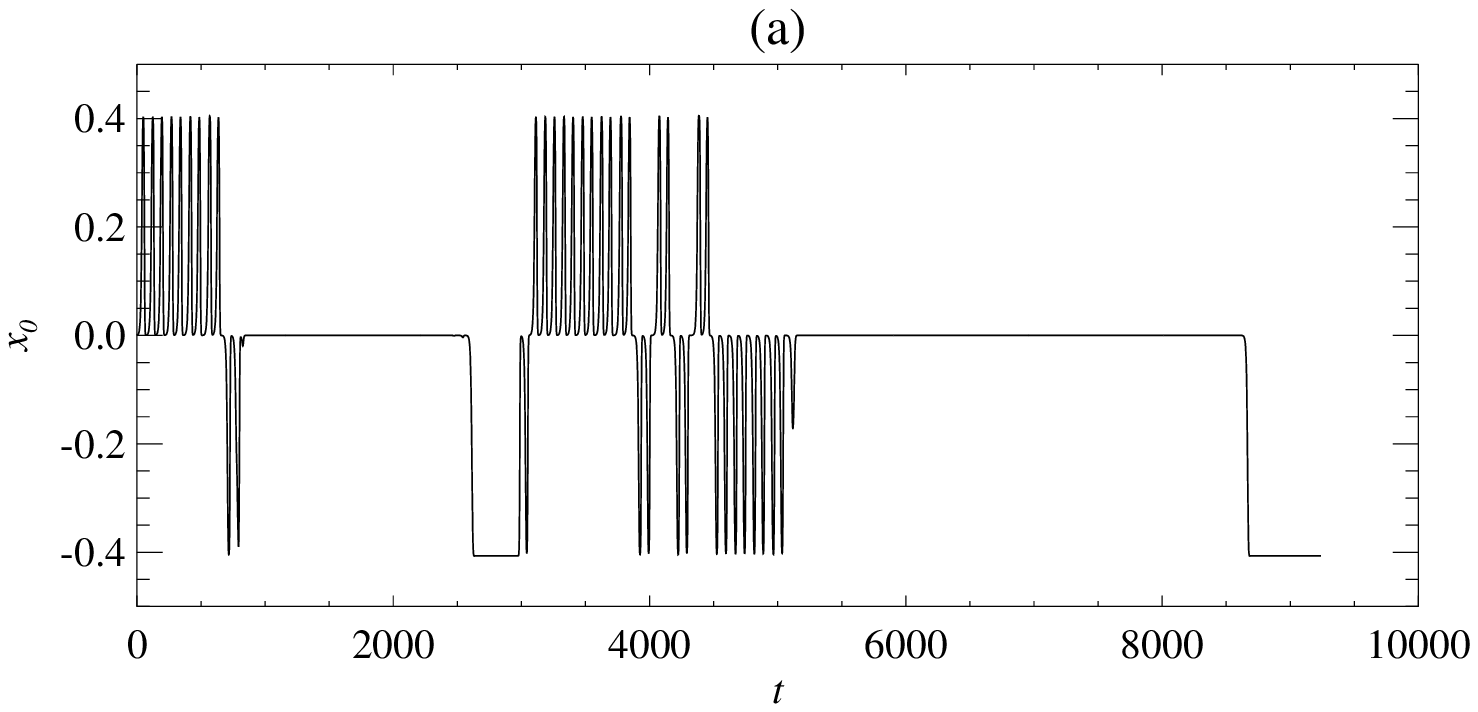,width=0.9\hsize}}
 \mbox{\psfig{file=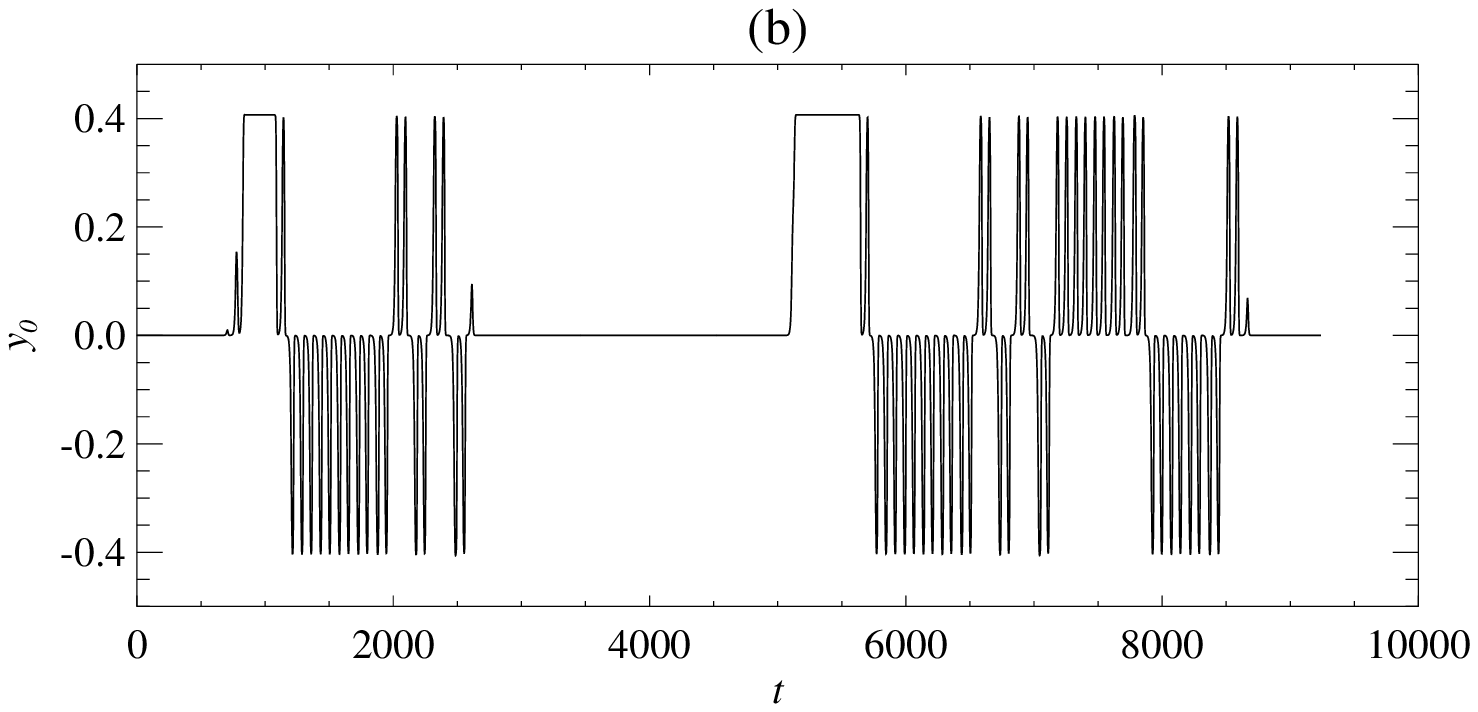,width=0.9\hsize}}
 \mbox{\psfig{file=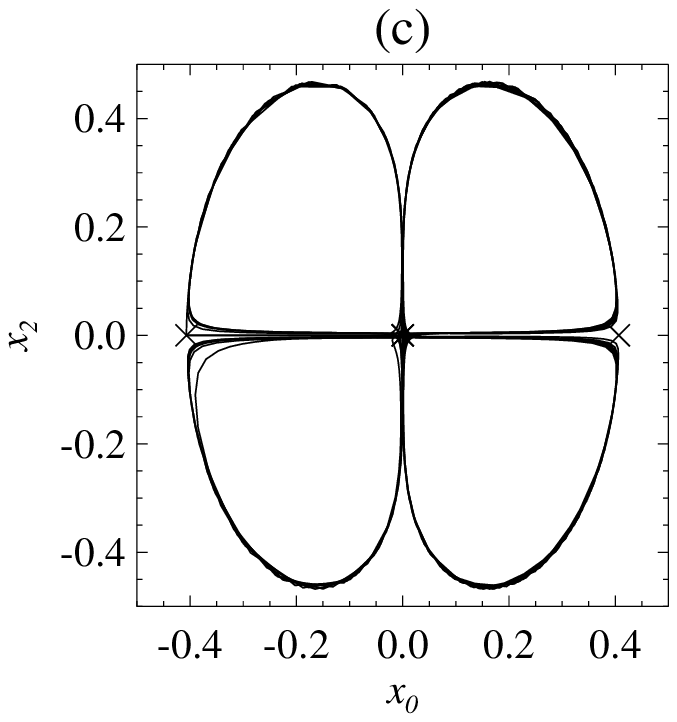,width=0.45\hsize}
       \psfig{file=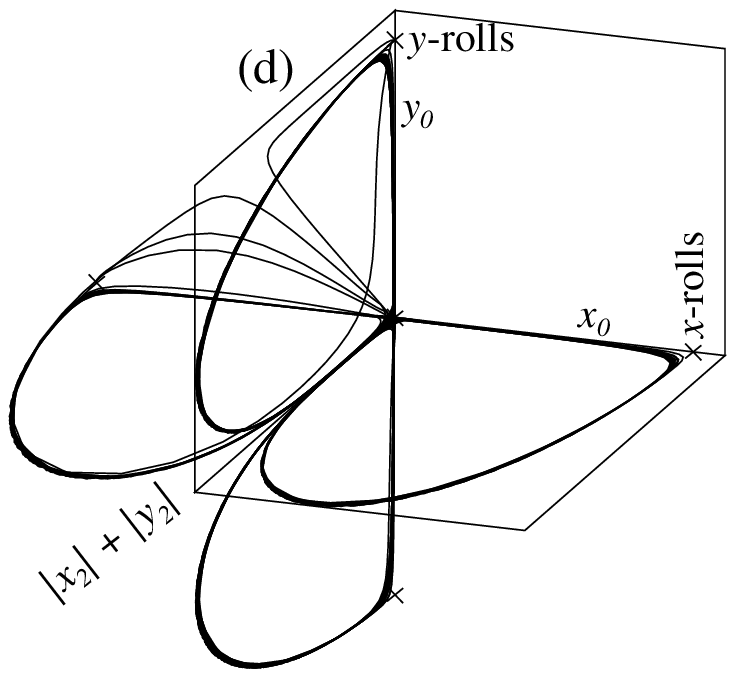,width=0.45\hsize}}
 \end{center}
 \caption{Numerical solutions of the model ODEs with $\mu=0.1655$, $Q=1/\pi^2$,
$\sigma=0.125$, $\zeta=0.05$ and $\beta=2.0$.
 (a)~$x_0$ against time; 
 (b)~$y_0$ against time; 
 (c)~$x_0$ against~$x_2$; 
 (d)~$x_0$, $y_0$ and $|x_2|+|y_2|$ in perspective.
 The crosses in (c) and (d) represent the $x$-roll, $y$-roll and trivial 
equilibrium points in $R_x$, $R_y$ and~$F$. Note that the chaotic attractor is
close to a heteroclinic cycle that connects the $x$-rolls, the $y$-rolls and 
the trivial fixed point.}
 \label{figphaseportraits}
 \end{figure}

In Figure~\ref{figphaseportraits} we show a typical example of timeseries when
there is attracting cycling chaos (with $\beta=2.0$). The system starts with
$x_0$ oscillating chaotically while $y_0$ is quiescent, switches to a state
where $y_0$ oscillates chaotically while  $x_0$ is quiescent, and so on. A more
careful examination reveals that after a switch the trajectory remains close to
a fixed point in $R_x$ or $R_y$ for an increasing length of time. Physically,
this corresponds to chaotic two-dimensional convection that switches between
rolls aligned in the $x$ and $y$ directions. Figure~\ref{figphaseportraits}(c)
shows the chaotic trajectories projected onto the $(x_0,x_2)$ plane, while (d)
illustrates switching between $P_x$ (the `horizontal' plane) and $P_y$ (the
`vertical' plane).

 \begin{figure}
 \begin{center}
 \mbox{\psfig{file=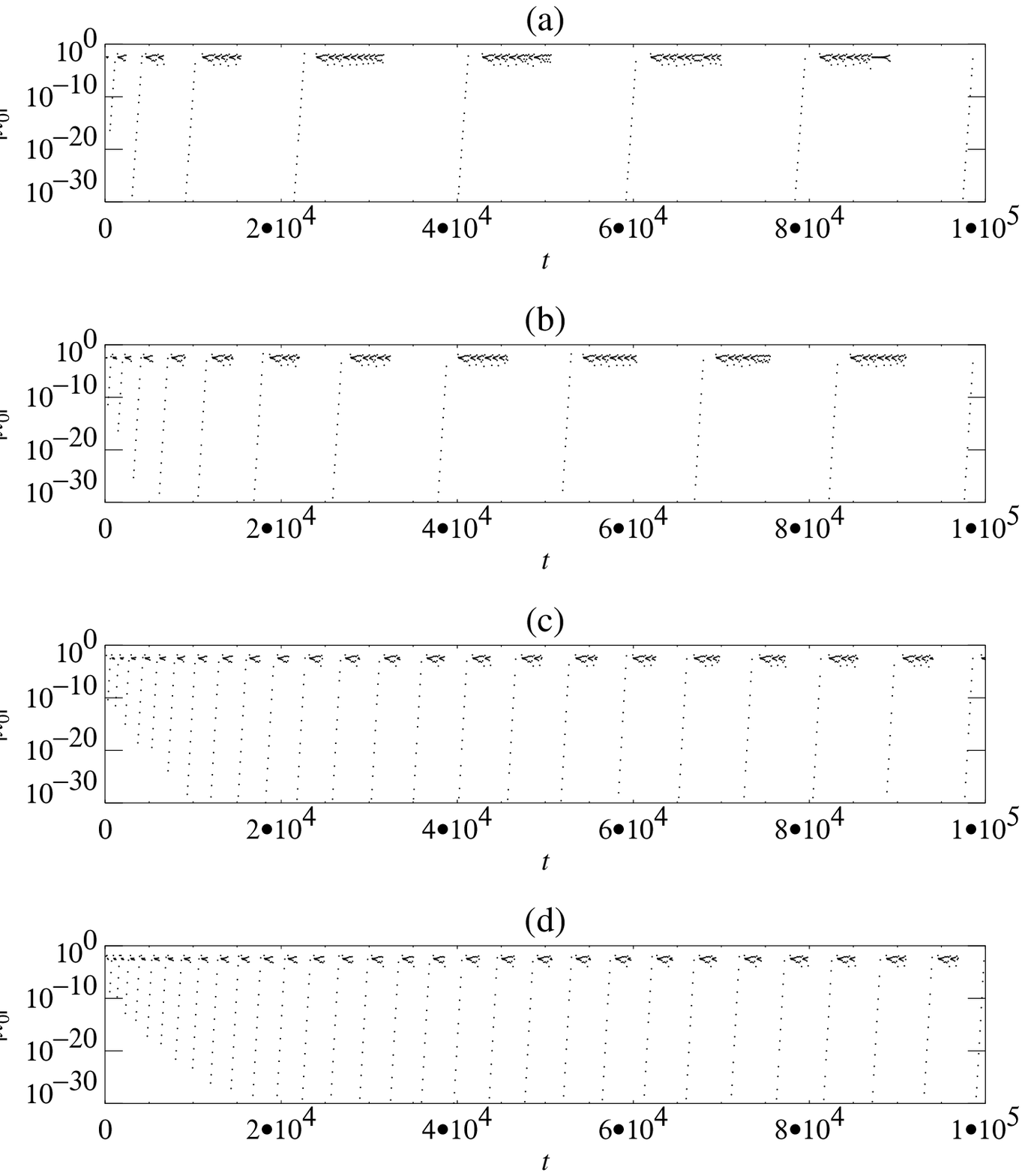,width=\hsize}}
 \end{center}
 \caption{Numerical solutions of the model ODEs: cycling chaos with
(a)~$\beta=2.00$, (b)~$\beta=1.80$, (c)~$\beta=1.65$ and (d)~$\beta=1.63$,
showing the values of $|x_0|$ at which the trajectory intersects the surface
defined by $|x_2|+|y_2|=0.01$. The rising exponential growth phases correspond
to the system switching from chaos in $P_y$ to chaos in $P_x$. The time between
switches increases as the system approaches the attracting cycling chaos, but 
the rate of increase depends on~$\beta$.}
 \label{figslowingdown}
 \end{figure}

Note how the chaotic behaviour in Figure~\ref{figphaseportraits}(a) and (b)
repeats: trajectories spend longer and longer near the unstable manifolds of
the $x$-roll and $y$-roll equilibrium points and take longer and longer between
each switch. This is illustrated further in Figure~\ref{figslowingdown}, which
shows intersections of a trajectory with the Poincar\'e surface
$|x_2|+|y_2|=0.01$ close to the trivial solution. There are two phases evident
in the cycle: the order one chaotic behaviour of $x_0$ near $P_x$ (while $y_0$
grows exponentially), and the exponential growth of $|x_0|$ as the trajectory
moves away from $P_y$ (while $y_0$ behaves chaotically). The time between
switches increases monotonically and the rate of increase varies with~$\beta$,
the normal parameter. In these numerical simulations, the switching time
saturates when certain components of the solution come close to the machine
accuracy of the computations (about $10^{-323}$ for double precision).

 \begin{figure}
 \begin{center}
 \mbox{\psfig{file=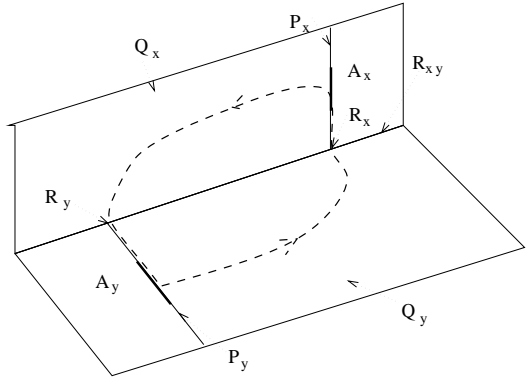,height=5cm}}
 \end{center}
 \caption{A schematic illustration of the location of the robust cycle relative
to the invariant subspaces forced by symmetry. The cycle
is between the chaotic invariant sets $A_x$ and $A_y$ (within $P_x$ and $P_y$)
and two fixed points contained in $R_x$ and $R_y$. The cycle is robust to
$G$-equivariant perturbations that fix the dynamics in $P_x$ and $P_y$. The 
intersection of $P_x$ and $P_y$ at the trivial solution is not shown, although 
$A_x$ and $A_y$ do in fact intersect there.}
 \label{figcycle}
 \end{figure}

We argue that this is evidence for attracting cycling chaos: trajectories
approach a structurally stable heteroclinic cycle between chaotic sets. In
Figure~\ref{figcycle} we show a schematic picture of the heteroclinic cycle. We
recall that the fixed point spaces $R_x$ and $R_y$ are 2 dimensional, $R_{xy}$
is 3 dimensional, $P_x$ and $P_y$ are 5 dimensional and $Q_x$ and $Q_y$ are 6
dimensional invariant subspaces in the 9 dimensional phase space. The system
starts near the $x$-roll equilibrium point in~$R_x$, which is unstable to shear
($x_2$); the parameters are such that the unstable manifold of $x$-rolls comes
close to the trivial solution and returns to a neighbourhood of $x$-rolls. This
global near-connection within $P_x$ is the source of the chaotic behaviour. We
refer to the chaotic sets in $P_x$ and $P_y$ as $A_x$ and $A_y$ respectively;
these contain the relevant roll equilibrium points and they both contain the
trivial solution, so there are structurally stable connections from the trivial
solution to the roll equilibrium points and from those to $A_x$ and $A_y$.
Within $Q_x$, $A_x$~is unstable to cross-rolls ($y_0$) since the trivial
solution is equally unstable in the $x_0$ and $y_0$ directions. Eventually
$y_0$ grows large enough that there is a switch to the $y$-roll equilibrium
point in~$R_y$, at which point $y_2$ starts to grow. Thus the cycle connects
invariant sets in the following fixed point subspaces:
 \begin{equation}
 \dots \rightarrow
 R_x \rightarrow
 P_x \rightarrow
 Q_x \rightarrow
 R_y \rightarrow
 P_y \rightarrow
 Q_y \rightarrow \dots
 \end{equation}
between the equilibrium points in $R_x$ and $R_y$, and $A_x$ and $A_y$ within
$P_x$ and~$P_y$, with the structurally stable connections needed to complete
the cycle from $A_x$ and $A_y$ to the $y$-roll and $x$-roll equilibrium points
lying within $Q_x$ and~$Q_y$.

Note that our scenario is certainly a simplification of the full set of
heteroclinic connections; there are other fixed points contained in the closure
of the smallest attracting invariant set, in particular the origin is contained
within the cycle.

\subsection{Blowout}
\label{secnumericsblowout}

 \begin{figure}
 \begin{center}
 \mbox{\psfig{file=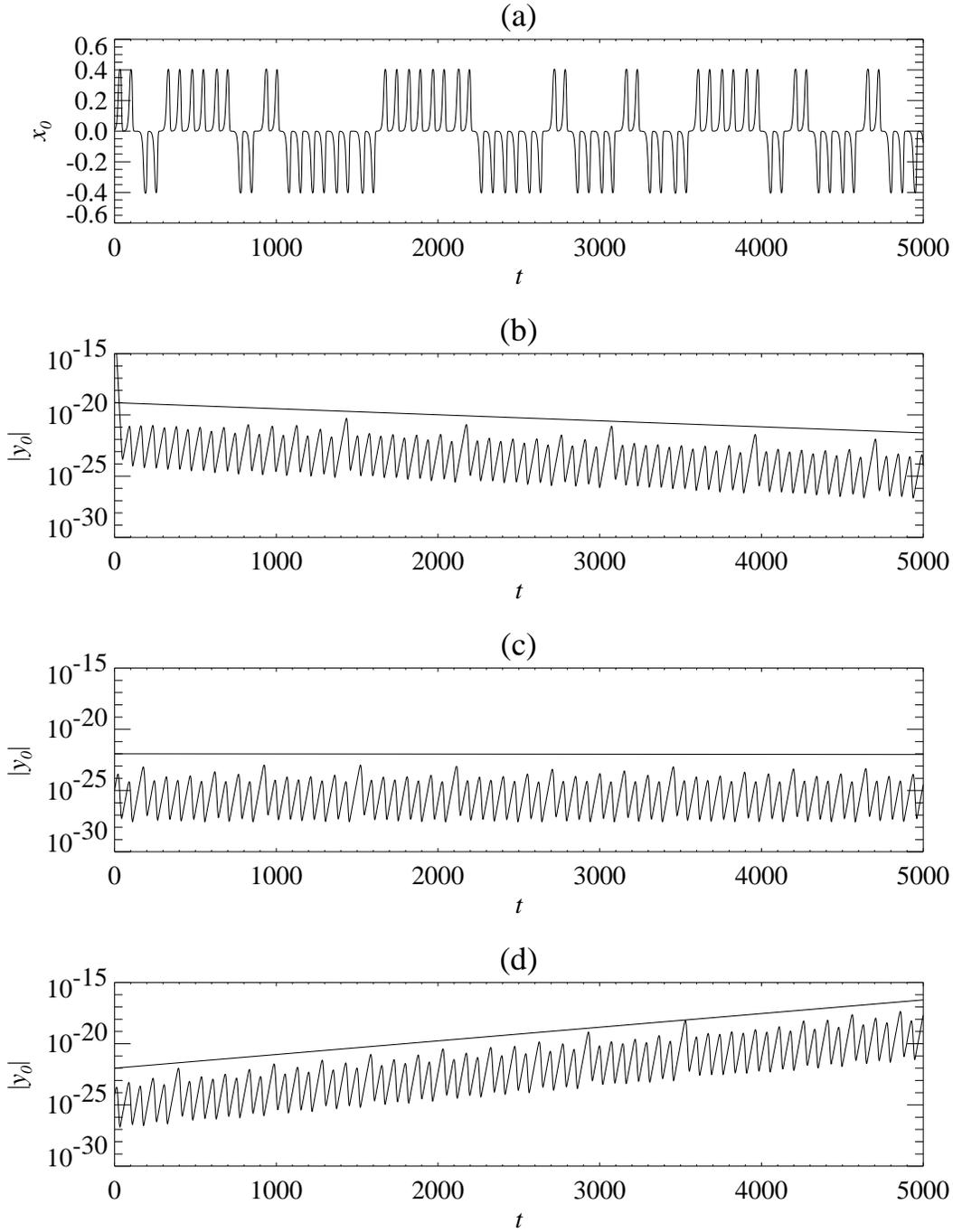,width=\hsize}}
 \end{center}
 \caption{Numerical solutions of the model ODEs. The top panel (a) shows the
chaotic time dependence of~$x_0$, independent of~$\beta$, and the lower three
panels show the blowout bifurcation: (b)~$\beta=3.50$, before the blowout
bifurcation (solutions decay to two-dimensional chaos); (c)~$\beta=3.47$ near
the blowout bifurcation; and (d)~$\beta=3.40$ after the blowout bifurcation
(solutions grow to cycling chaos). The straight lines show the average growth
or decay rates predicted by the Liapunov exponents calculated 
using~({\protect\ref{eqODELiapunov}}).}
 \label{figblowout}
 \end{figure}

We turn now to the question of how the cycling chaos is created. With
$\beta=3.5$ we find attracting two-dimensional chaos
(Figure~\ref{figblowout}a,~b), which loses stability around $\beta=3.47$~(c) in
a blowout bifurcation, and for $\beta=3.40$~(d), there is exponential growth
away from~$P_x$ into~$Q_x$. Within $Q_x$, the $y$-roll equilibrium points are
sinks, establishing the structurally stable connection from the chaos in $P_x$
to the equilibrium point in~$R_y$, and hence the creation of cycling chaos.

\subsection{Resonance}
\label{secnumericsresonance}

 \begin{figure}
 \begin{center}
 \mbox{\psfig{file=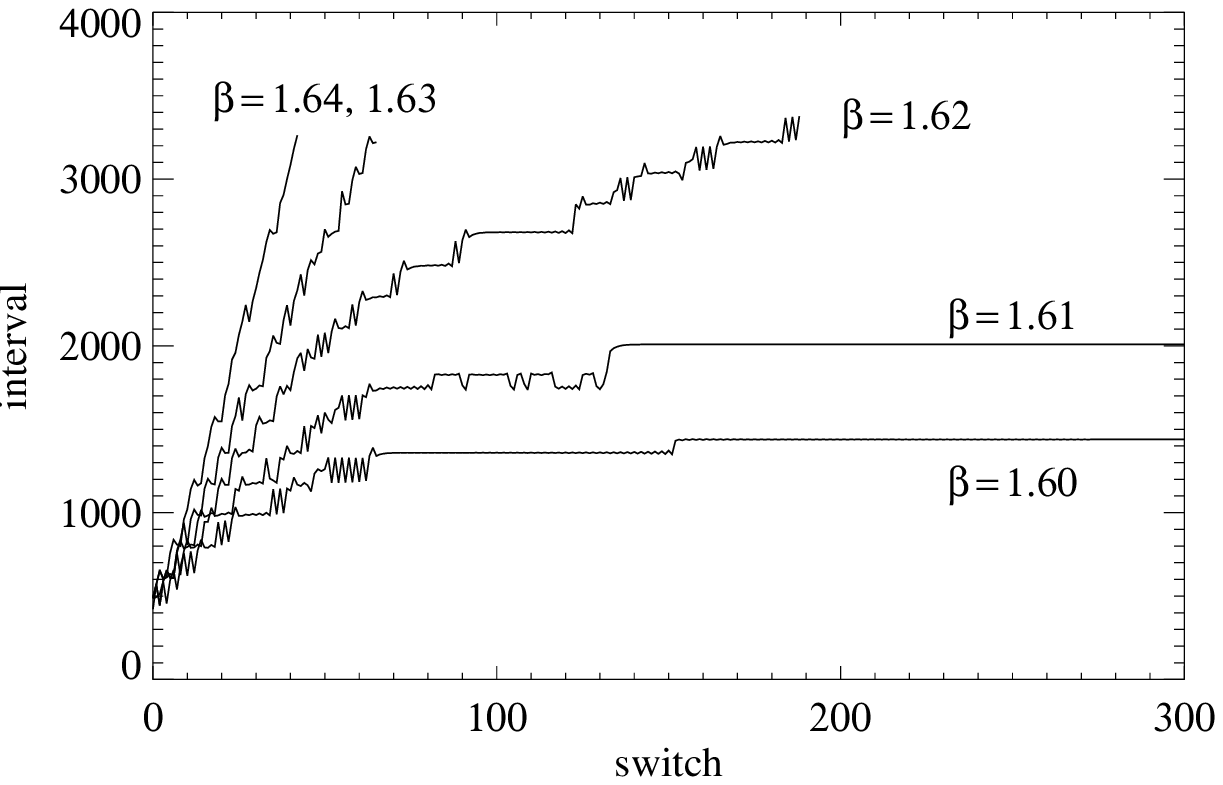,width=\hsize}}
 \end{center}
 \caption{Time intervals between switches between $P_x$ and $P_y$
with $\beta$ in the range $1.60$--$1.64$. The intervals between switches
increase by a factor of about $1.03$ and $1.02$ per switch for $\beta=1.64$ and
$1.63$ respectively. The switching times stop increasing once the values of
variables become so small that the cubic terms in the ODEs cannot be
represented accurately with double-precision numbers (and calculations
were terminated once any variable went below $10^{-100}$). For $\beta=1.61$ and
$1.60$, the system approaches a periodic orbit and the intervals between
switches goes to a constant (though in the case of $\beta=1.61$, the periodic
orbit is only achieved after 1500 switches).}
 \label{figresonance}
 \end{figure}

As illustrated in Figures~\ref{figslowingdown}, as $\beta$ decreases
towards~$1.63$, trajectories spend progressively longer in each visit to $P_x$
and $P_y$ before switching to the conjugate chaotic invariant set. Eventually,
trajectories come arbitrarily close to the invariant subspaces $P_x$ and $P_y$
(limited only by machine accuracy in the numerical simulations).
Figure~\ref{figresonance} shows how the time intervals between switches between
$A_x$ and $A_y$ increases as the system approaches this heteroclinic cycle, and
how the rate of approach to the heteroclinic cycle decreases as $\beta$
approaches~$1.62$. The intervals between switches grow by a factor of about
$1.4$, $1.2$ and $1.1$ per switch for $\beta=2.00$, $\beta=1.80$ and
$\beta=1.65$ respectively, and $1.03$ and $1.02$ for $\beta=1.64$ and $1.63$.
By $\beta\approx1.62$, the heteroclinic cycle is no longer attracting, and for
$\beta=1.61$ and $1.60$, the system settles down to periodic behaviour that is
bounded away from $P_x$ and $P_y$, though the periodic orbits are actually
quite close to these invariant subspaces. For these calculations, we imposed a
cut-off of~$10^{-100}$: the calculation ceased once any variable became smaller
than this.

In the next section, we derive a map that allows us to compute longer
trajectories more accurately, and we demonstrate cycling chaos, its creation in
a blowout bifurcation and its loss of attractiveness using this map. We analyse
the blowout bifurcation in section~\ref{secblowout}, and argue in
section~\ref{secresonance} that the cycling chaos created in that bifurcation
ceases to be attractive at a resonance.

\section{Reduction to a map}
\label{secmap}

In this section, we discuss a map that models the behaviour of the ODEs in the
parameter regime described above, and that helps clarify what happens near
blowout bifurcation and the resonance of the cycling chaos.

\subsection{Derivation}

We rely on a map derived for the two-dimensional dynamics by breaking up the
flow into pieces near and between equilibrium points~\cite{RM96}. Within the
$P_x$ subspace, the leading stable eigendirection of the origin (that is, the
eigendirection with negative eigenvalue closest to zero) is in the $(x_2,a)$
plane, and the unstable direction of the origin is along the $x_0$~axis. All
trajectories leaving the origin in that direction follow the structurally
stable connection within $R_x$ to one or other of the $x$-roll equilibrium
points. The one-dimensional unstable manifold of $x$-rolls lies within $P_x$,
and the chaotic attractor is associated with a global bifurcation at which that
unstable manifold collides with the origin. Near this global bifurcation, the
dynamics within $P_x$ is modelled by an augmented Lorenz map:
 \begin{equation}\label{eqlorenz}
 (x_0,x_2=\pm1)\mapsto
 \left(\sgn(x_0)(-\kappa+C_1|x_0|^{\delta_1}),-x_2\right)
 \end{equation}
defined as a map from the surface of section $|x_2|=h$ to itself. Details of
the derivation are given in~\cite{RM96}, but briefly, $\kappa$ is a parameter
related to $\mu$ in the ODEs, with $\kappa=0$ at the global bifurcation; $h$~is
a small positive constant that we scale to one; $\sgn(x)=+1$ if $x>0$ and $-1$
if $x<0$; $\delta_1$ depends on the ratio of various stable and unstable
eigenvalues at the origin and the $x$-roll equilibrium points (with
$0<\delta_1<1$); and $C_1$ is a (negative) constant.

It is a straightforward matter to include the effect of a small perturbation in
the $y_0$ and $y_2$ directions. Near the $P_x$ subspace, $y_0$ and $y_2$ will
grow linearly at a rate that depends on~$x_0$. This means that we get a mapping
of the form
 \begin{eqnarray}\label{eqnearPx}
 \lefteqn{(x_0,x_2=\pm1,y_0,y_2)\mapsto{}} \nonumber\\
 &&\left(\sgn(x_0)(-\kappa+C_1|x_0|^{\delta_1}),-x_2,
    C_2 y_0 |x_0|^{\delta_2}, C_3 y_2 |x_0|^{\delta_3}\right),
 \end{eqnarray}
where $C_2$ and $C_3$ are constants and the exponents $\delta_2$ and $\delta_3$
again depend on ratios of eigenvalues, with $\delta_2<0$ and $\delta_3>0$. The
exponent $\delta_2$ is negative since a small value of $x_0$ implies that the
trajectory spends longer near the origin, so $y_0$ has a longer time to grow.
Similarly, near~$P_y$ we get the mapping
 \begin{eqnarray}\label{eqnearPy}
 \lefteqn{(x_0,x_2,y_0,y_2=\pm1)\mapsto{}} \nonumber\\
 &&\left(C_2x_0|y_0|^{\delta_2}, C_3x_2|y_0|^{\delta_3},
         \sgn(y_0)(-\kappa+C_1|y_0|^{\delta_1}), -y_2\right),
 \end{eqnarray}
as long as $x_0$ is much smaller than~$y_0$.

These maps are valid provided that the trajectory remains close to the $P_x$ or
$P_y$ subspaces. We model the switch from behaviour near $P_x$ to near $P_y$ by
assuming that (\ref{eqnearPx}) holds provided that $|x_0|>|y_0|$; otherwise,
the trajectory leaves the neighbourhood of the origin along the $y_0$-axis,
visits a $y$-roll equilibrium point and returns to the surface of section
$|y_2|=1$ near the origin:
 \begin{eqnarray}\label{eqswitchPxPy}
 (x_0,x_2=\pm1,y_0,y_2)&\mapsto&
   \left(C_4 x_0 |y_2|^{\delta_4}|y_0|^{\delta_2}, 
       \pm C_5 |y_2|^{\delta_5}|y_0|^{\delta_3},\right.\\
 &&\phantom{\left(\right.\kern-2\nulldelimiterspace}
     \left.
     \sgn(y_0)(-\kappa+C_6 |y_2|^{\delta_6}|y_0|^{\delta_1}),
       \sgn(y_2)\right),\nonumber
 \end{eqnarray}
where, as above, $C_4$, $C_5$ and $C_6$ are constants and $\delta_4$,
$\delta_5$ and $\delta_6$ are ratios of eigenvalues. Similarly,
(\ref{eqnearPy}) holds if $|x_0|<|y_0|$; otherwise the trajectory switches from
$P_y$ to~$P_x$:
 \begin{eqnarray}\label{eqswitchPyPx}
 (x_0,x_2,y_0,y_2=\pm1)&\mapsto{}&
   \left(\sgn(x_0)(-\kappa+C_6|x_2|^{\delta_6}|x_0|^{\delta_1}),
        \sgn(x_2),\right.\\
 &&\phantom{\left(\right.\kern-2\nulldelimiterspace}
     \left.
       C_4y_0|x_2|^{\delta_4}|x_0|^{\delta_2},
       \pm C_5|x_2|^{\delta_5}|x_0|^{\delta_3}\right).\nonumber
 \end{eqnarray}
 Then in the full map, as long as $|x_0|>|y_0|$, the trajectory behaves
chaotically under~(\ref{eqnearPx}) (near the $P_x$ subspace), with $x_0$
obeying a Lorenz map and $y_0$ growing or decaying according to the value
of~$x_0$. If $y_0$ grows sufficiently that $|y_0|>|x_0|$, there is one iterate
of map~(\ref{eqswitchPxPy}), which makes~$x_0$ small as the system switches
from $P_x$ to $P_y$, followed by many iterates of~(\ref{eqnearPy}), and one
iterate of~(\ref{eqswitchPyPx}) as the system switches back to~$P_x$.

The $\delta$~exponents can be determined from the eigenvalues of the trivial
solution and of $x$-rolls. The relevant eigenvalues of the origin are (in the
notation of \cite{RM96}) the growth rate~$\mu$ of $x_0$ and $y_0$ and the
slowest decay rate $\lambda_\zeta$ of $(x_2,a)$, determined by the eigenvalue
of
 \begin{equation}
 \left[\matrix{-\sigma&-\sigma Q\cr
               \zeta&-\zeta\cr}\right]
 \end{equation}
closest to zero. The relevant eigenvalues of $x$-rolls are the growth
rate~$\lambda^+$ and the slowest decay rate $\lambda_2^-$ of $(x_1,x_2,a)$, 
determined by the eigenvalues of
 \begin{equation}
 \left[\matrix{-\nu&\sqrt{\mu}&0\cr
               \gamma\sqrt{\mu}&-\sigma&-\sigma Q\cr
               0&\zeta&-\zeta\cr}\right].
 \end{equation}
The other important eigenvalues at the $x$-roll equilibrium point are the decay
rate of $y_0$ ($-\beta\mu$) and the decay rate of $y_2$ ($\lambda_\zeta$). Then
we have
 \begin{eqnarray}\label{eqndeltas}
 \delta_1&=& \frac{\displaystyle-\lambda_\zeta}
                  {\displaystyle\mu}\left(1+\delta_6\right),
 \qquad
 \delta_2 =  \frac{\displaystyle-\lambda_\zeta}
                  {\displaystyle\mu}\delta_4 - 1,
 \qquad
 \delta_3 =  \frac{\displaystyle-\lambda_\zeta}
                  {\displaystyle\mu}\left(1+\delta_5\right),    \nonumber \\
 \delta_4&=& \frac{\displaystyle\beta\mu}
                  {\displaystyle\lambda^+},
 \qquad
 \delta_5 =  \frac{\displaystyle-\lambda_\zeta}
                  {\displaystyle\lambda^+},                           
 \qquad
 \delta_6 =  \frac{\displaystyle-\lambda_2^-}
                  {\displaystyle\lambda^+}.                     
 \end{eqnarray}
 From fitting the map to trajectories of the ODEs with $\beta=3.47$ (see 
Figure~\ref{figblowout}b), we find values for the constants:
 \begin{eqnarray}\label{eqnconsts}
 \kappa&=&-0.014072,      \nonumber\\
 C_1=-0.12450, \qquad C_2 &=& 0.27200, \qquad C_3 = 0.16791,   \\
 C_4=609.65,   \qquad C_5 &=& 0.86515, \qquad C_6 = 0.0015974.  \nonumber
 \end{eqnarray}
The eigenvalues that do not depend on $\beta$ are:
 \begin{equation}\label{eqneigenvalues}
 \lambda_\zeta=-0.059697,  \qquad
 \lambda^+    = 0.29606,   \qquad
 \lambda_2^-  =-0.048978.  
 \end{equation}
With these eigenvalues, we have $\delta_1=0.42038$, $\delta_3=0.43344$,
$\delta_5=0.20164$ and $\delta_6=0.16543$, while $\delta_2$ (negative) and
$\delta_4$ depend on~$\beta$.

\subsection{Iterating the map}

 \begin{figure}
 \begin{center}
 \mbox{\psfig{file=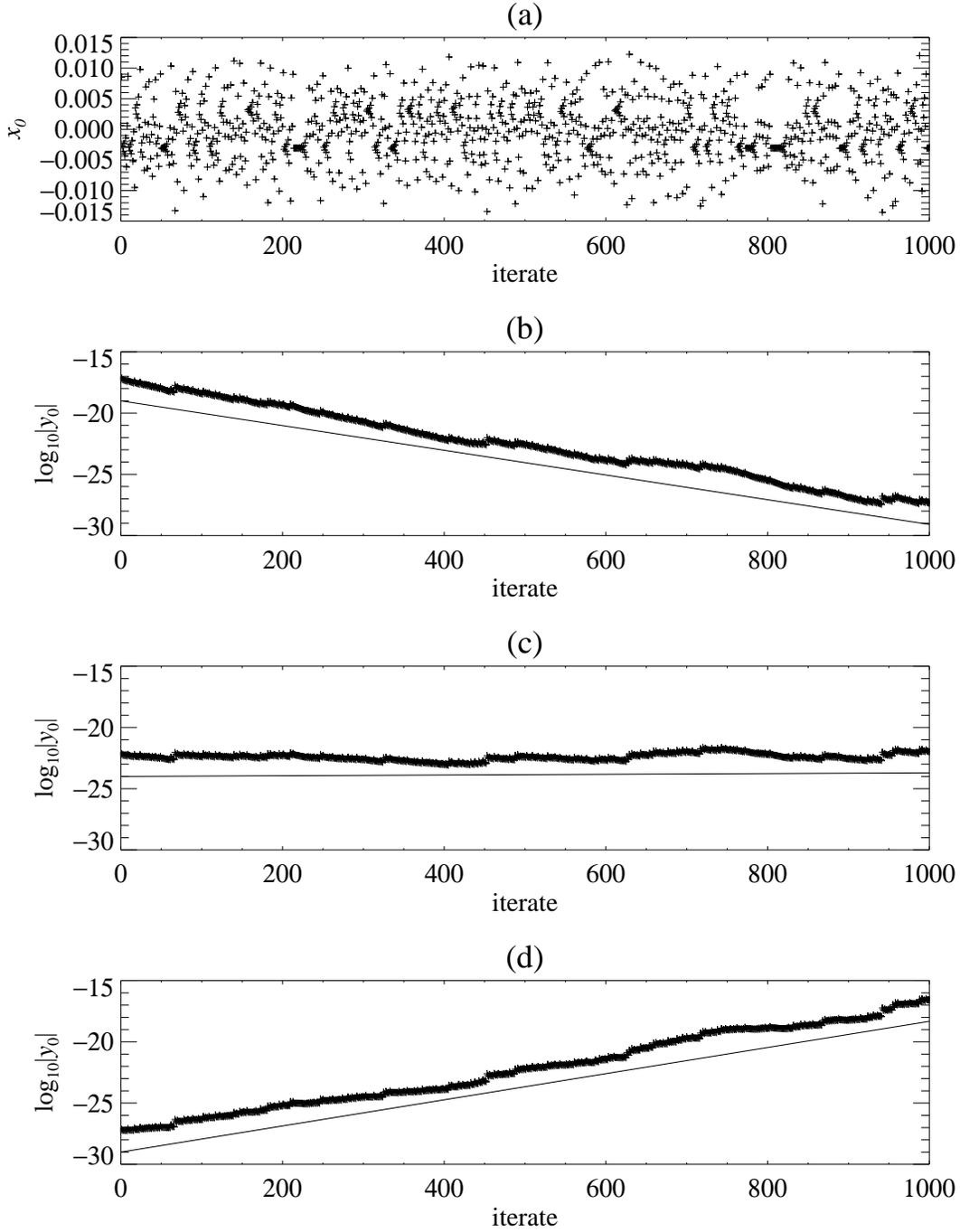,width=\hsize}}
 \end{center}
 \caption{Blowout in the map~({\protect\ref{eqnearPx}--\ref{eqswitchPyPx}}):
compare with Figure~{\protect\ref{figblowout}}.
 (a)~Chaotic behaviour of $x_0$ near $P_x$. For different values 
of~$\beta$ (b:~$\beta=3.89$, c:~$\beta=3.87$, d:~$\beta=3.85$), the 
distance from $P_x$ as measured by~$|y_0|$ decays in~(b) and grows in~(d). 
The blowout bifurcation occurs for parameters near~(c). The straight lines show 
the average growth or decay as predicted by~(\ref{eqlambdae}), averaged over 
1000 iterates.}
 \label{figmapone}
 \end{figure}

We seek to reproduce in the map what we have observed in the ODEs: the blowout
bifurcation that creates cycling chaos, and the loss of attractiveness of
cycling chaos. The first of these (Figure~\ref{figmapone}) is straightforward:
typical Lorenz-like chaos is shown in Figure~\ref{figmapone}(a), while the
change from decay towards $P_x$ to growth away from~$P_x$ at $\beta\approx3.87$
is shown in (b--d).

 \begin{figure}
 \begin{center}
 \mbox{\psfig{file=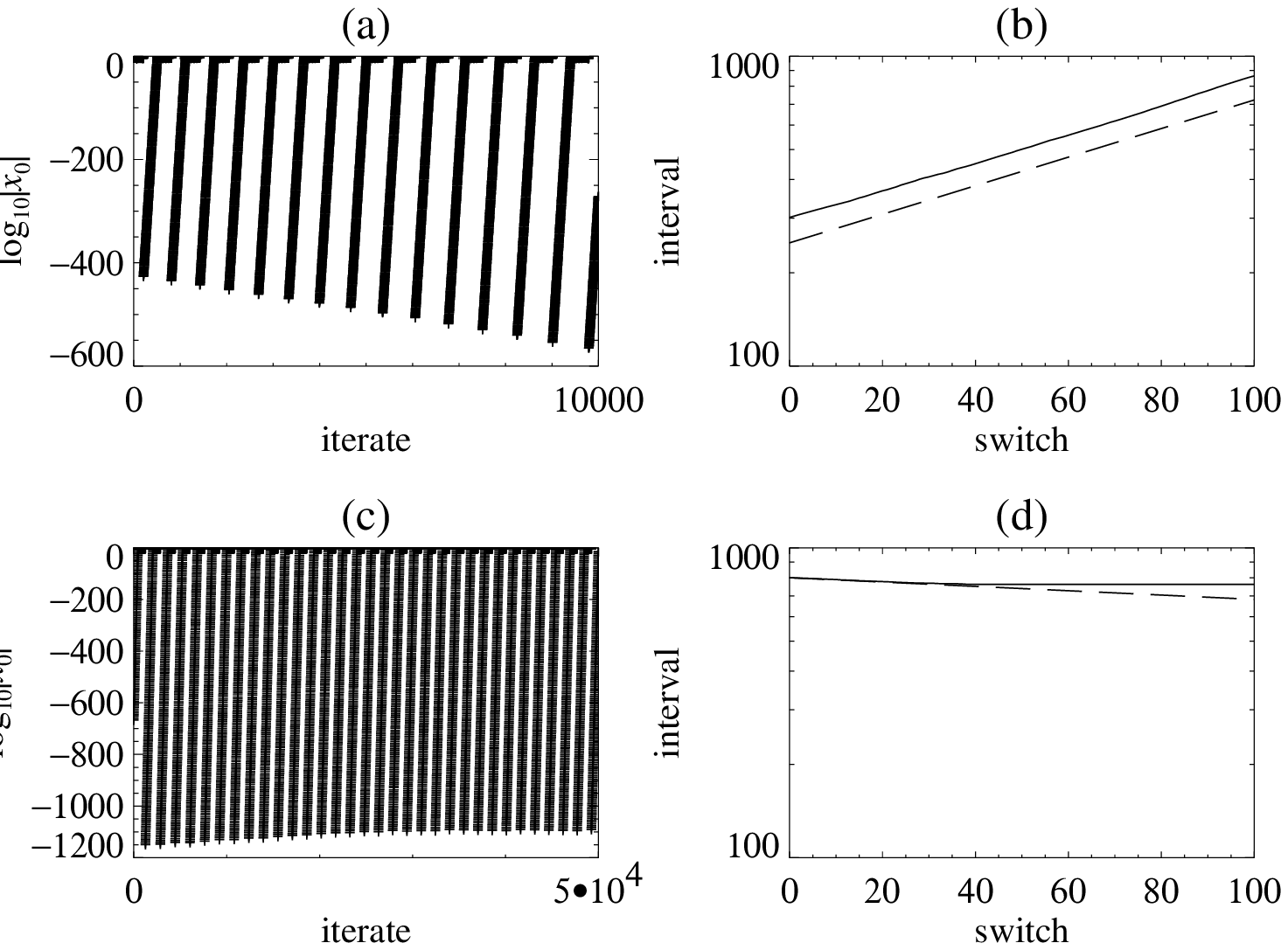,width=\hsize}}
 \end{center}
 \caption{Resonance in the map~({\protect\ref{eqnearPx}--\ref{eqswitchPyPx}}):
compare with Figures~{\protect\ref{figslowingdown}}
and~{\protect\ref{figresonance}}. (a) and~(b): $\beta=1.100$, showing the
approach to cycling chaos and the increase in number of iterates between
switches ($\rho=1.011$). (c) and~(d): $\beta=1.088$, showing growth away from
cycling chaos ($\rho=0.9984$) and saturation to a periodic orbit. The dashed
lines in (c) and (d) show the predicted rate of increase or decrease of the
switching times from~({\protect\ref{eqrhomap}}).}
 \label{figmaptwo}
 \end{figure}

Cycling chaos is found after the blowout bifurcation (Figure~\ref{figmaptwo}a):
the system switches between chaos in $P_x$ and chaos in $P_y$, getting closer
and closer to the invariant subspaces after each switch and spending longer
between switches (Figure~\ref{figmaptwo}b). The rate of increase of the
intervals between switches depends on~$\beta$, and is about a factor of 1.01
per switch for $\beta=1.10$. Decreasing~$\beta$ to $1.088$
(Figure~\ref{figmaptwo}c,d) results in growth away from cycling chaos and
saturation to a periodic orbit.

These calculations required formulating the map
(\ref{eqnearPx}--\ref{eqswitchPyPx}) in terms of the logarithms of the
variables in order to cope with the large ($10^{1000}$) dynamic range. The one
place in which accuracy is inevitably lost is in the switch from $P_x$ to $P_y$
(and back), using~(\ref{eqswitchPxPy}): the
$C_6|y_2|^{\delta_6}|y_0|^{\delta_1}$ term is swamped by the order
one~$\kappa$. As a result of this, the chaotic trajectories start in exactly
the same way each time the system switches from one invariant subspace to the
other.

In terms of the ODEs, the trajectory entering a neighbourhood of $P_x$ close to
$x$-rolls shadows the unstable manifold of that equilibrium point (lying
inside~$P_x$ and leading to the chaotic set~$A_x$) until the switch to~$P_y$.
This is in agreement with the ODE behaviour shown in
Figure~\ref{figphaseportraits}.

\section{Analysis of the blowout bifurcation}
\label{secblowout}

We briefly review some definitions. If $A$ is a compact flow-invariant subset
then we define the unstable set
 \begin{equation}
 \cW^u(A)=\{ x\in\Rset^9~:~\alpha(x)\subseteq A\}
 \end{equation}
and the stable set (or {\em basin of attraction\/})
 \begin{equation}
 \cW^s(A)=\{ x\in\Rset^9~:~\omega(x)\subseteq A\}
 \end{equation}
where $\alpha(x)$ (resp. $\omega(x)$) is the limit set of a trajectory of the
ODE passing through $x$ in the limit $t\rightarrow-\infty$ (resp. $\infty$). We
say a compact invariant set $A$ is an {\em attractor\/} in the sense of Milnor
if
 \begin{equation}\label{eqMattr}
 \ell(\cW^s(A))>0
 \end{equation}
where $\ell(\cdot)$ is Lebesgue measure on $\Rset^9$. It is said to be a {\em
minimal Milnor attractor\/} if there are no proper compact invariant subsets
$S\subset A$ with $\ell(\cW^s(A)\setminus\cW^s(S))>0$ \cite{Mil85}.

As shown in \cite{Ale&al92}, Milnor attractors can occur in a robust manner if
the attractor lies within an invariant subspace. Suppose $P$ is an invariant
subspace and $A$ is a compact invariant set in $P$ such that $A$ is a minimal
Milnor attractor for the flow restricted to $P$ and such that $A$ has a natural
ergodic invariant measure $\mu$ for the flow restricted to $P$. It is possible
to show (under certain additional hypotheses \cite{Ale&al92}) that $A$ as an
attractor in the full system if and only if $\Lambda(\mu)<0$, where $\Lambda$
is the most positive Liapunov exponent for $\mu$ in a direction transverse to
$P$ (see \cite{AshBueSte96} for a detailed discussion). If we have access to a
normal parameter \cite{AshBueSte96} such that we can vary the normal dynamics
without changing the dynamics in~$P$, we can vary $\Lambda$ through zero and
observe loss of attractiveness of $A$ at what Ott and Sommerer have termed a
{\em blowout bifurcation\/} \cite{Ott&Som94}.

\subsection{Set criticality of a blowout bifurcation}

Ott and Sommerer identify two possible scenarios at blowout. At a {\em
supercritical blowout\/} the attractor bifurcates to a family of attractors
displaying on-off intermittency \cite{Pla&al93}, with trajectories that come
arbitrarily close to $A$ and thus linger near $A$ for long times (but with a
well-defined mean length of lingering or `laminar phase'). At {\em subcritical
blowout\/} there are no nearby attractors after loss of stability of $A$. Note
that \cite{AshAstNic96} discusses the question of how to distinguish these
cases.

However, what we see in this paper is that there is at least one other
possibility at blowout bifurcation that is also robust to normal perturbations,
namely a bifurcation to a cycling state or a robust heteroclinic cycle between
chaotic invariant sets. This can be seen to be set supercritical but not
supercritical, in the following sense.

By reparametrising if necessary we can assume that we have a normal parameter
 \begin{equation}
 c=\Lambda(\mu)
 \end{equation}
so $A$ undergoes a blowout bifurcation at $c=0$. By the argument above, $A$ is
an attractor only if $c<0$.

If there is a family of minimal Milnor attractors $A_c$, $c>0$ such that for
all $\epsilon>0$
 \begin{equation}
 \overline{\bigcup_{c\in(0,\epsilon)} A_c} \supseteq A
 \end{equation}
then we say that the blowout is {\em set supercritical}. Note that as
discussed in \cite{Ott&Som94} and implied by the very word `blowout', we cannot
typically expect the limit of the attractors to just be the set $A$.

If in addition $A_c$ supports a family of natural ergodic invariant measures
$\mu_c$ ($c>0$) and $\mu_c \rightarrow \mu$ as $c\rightarrow 0$ then we say the
blowout is {\em measure supercritical\/} or just {\em supercritical}.
(Convergence is the in the $C^*$ topology on probability measures.)  This
definition of criticality was used in \cite{AshAstNic96}.

If the blowout is not set supercritical then we say the blowout is {\em
subcritical}.

It seems that one will often get bifurcations in symmetric systems that are set
supercritical but not supercritical. For example one can generically get a
blowout from a group orbit of attractors (under a finite group) which yields a
single attractor that limits onto the whole group orbit. Moreover, in the
cycling chaos studied here, the attractor after blowout includes not only the
chaotic invariant sets, but also fixed points involved in the heteroclinic
cycle. As it is a cycle, it does not possess a natural ergodic invariant
measure \cite{Sig92} and averages of observables from the system in this state
will typically not converge but continue to oscillate more and more slowly.

The blowout scenario described above holds only for variation of normal
parameters; in general the variation of a parameter will affect both the normal
dynamics and the dynamics within the invariant subspace. If the dynamics in the
invariant subspace is chaotic, we can expect to see a large number of
bifurcations happening within the invariant subspace and these will cause the
blowout to be spread over an interval of parameter values; there is no reason
why $\Lambda(\mu_x)$ should vary continuously with a parameter that varies
$\mu_x$ in a very discontinuous manner.

Nevertheless, for the numerical results presented in Section~\ref{secnumerics}
the dynamics in the invariant subspace vary in quite regular way. This is
because in our system the parameter $\beta$ is a normal parameter for the
attractor $A_x$; for more discussion of normal parameters, see
\cite{AshBueSte96,Cov&al97}.

\subsection{Evidence of blowout bifurcation in the ODE model}

By mechanisms described in \cite{RM96} the pure $x$ and $y$ chaotic dynamics
corresponding to dynamics within $P_x$ and $P_y$ respectively can become
chaotic by means of a symmetric global bifurcation that generates Lorenz-like
attractors approaching the equilibrium solution with full symmetry and $x$-roll
or $y$-roll equilibrium solutions in $R_x$ or~$R_y$.

Now suppose there exist chaotic attractors $A_x$ and $A_y$ contained in $P_x$
and $P_y$ (on average they have more symmetry), From here on we will mostly
discuss $A_x$ but note that as $A_y$ is a conjugate attractor, the same will
hold for~$A_y$.

These attractors contain a saddle equilibrium $e$ in $F$ (the trivial solution)
and so in particular they cannot be Liapunov stable attractors. This is because
$W^u(e)\cap P_x$ must be non-trivial manifold (otherwise $e$ is an attractor;
thus $W^u\cap P_y$ is also non-trivial as $\iso(P_y)$ is conjugate to
$\iso(P_x)$.  Therefore
 \begin{equation}
 \cW^u(A_x)\not\subseteq A_x
 \end{equation}
and so $A_x$ cannot be Liapunov stable. However it is possible that $A_x$ is an
attractor in the sense of Milnor; this will imply that the basin will be
locally riddled in the sense of \cite{AshBueSte96}.

We assume that $A_x$ and $A_y$ are minimal Milnor attractors containing $e$. We
also assume that they have natural ergodic invariant measures $\mu_x$ and
$\mu_y$ supported on them. We now concentrate on $A_x$. Whether $A_x$ is an
attractor depends on its the spectrum of normal Liapunov exponents. Note that
the zero Liapunov exponent corresponding to time translation always corresponds
to a perturbation tangential to the invariant subspace. If
 \begin{equation}
 \Lambda(\mu_x)<0,
 \end{equation}
where $\Lambda(\mu)$ is the most positive normal Liapunov exponent for the
measure~$\mu$, it is possible to show that $A_x$ satisfies~(\ref{eqMattr})
\cite{AshBueSte96}, and hence that $A_x$ is a Milnor attractor.

It is comparatively easy to compute $\Lambda(\mu_x)$ in that case, as the
largest transverse Liapunov exponent of $A_x$ in the parameter regime discussed
corresponds to directions in the $Q_x$ direction, where linearised
perturbations are described by
 \begin{equation}
 \dot{y}_0=\left(\mu+\theta(t)-\beta x_0(t)^2\right)y_0.
 \end{equation}
Thus we can see that
 \begin{equation}
 \Lambda(\mu_x)=\mu + \langle \theta \rangle_{\mu_x}
                    - \beta \langle x_0^2\rangle_{\mu_x},
 \end{equation}
where $\langle f(x)\rangle_\mu=\int f(x)d\mu(x)$. From the equation for
$\dot{\theta}$ it is possible to show that
 \begin{equation}
 \lim_{T\rightarrow\infty}\frac{1}{T}\int^T \theta\,dt=
 \lim_{T\rightarrow\infty}\frac{1}{T}\int^T x_0^2 \,dt
 \end{equation}
along bounded trajectories of the ODE. For the given parameter values it is
possible to approximate $\langle \theta \rangle_{\mu_x}=-0.03703$ and so
$\langle x_0^2\rangle_{\mu_x}=0.03703$. Thus
 \begin{equation}\label{eqODELiapunov}
 \Lambda(\mu_x) 
  = \mu - \langle x_0^2\rangle_{\mu_x} - \beta \langle x_0^2\rangle_{\mu_x}
  = 0.12847-0.03703\beta,
 \end{equation}
implying that the blowout bifurcation occurs at approximately $\beta=3.47$,
which is in good agreement with the simulations (Figure~\ref{figblowout}b).

Note that whenever $e$ is hyperbolic and within the attractor $A_x$ there
exists at least one ergodic invariant measure $\mu_e$ (in particular that one
supported on~$e$) such that $\Lambda(\mu_e)>0$. In particular, this means that
the basins of the attractors $A_x$ and $A_y$ are riddled for all parameter
values in our problem!

Likewise, in the map~(\ref{eqnearPx}) near~$P_x$, the logarithm of~$y_0$ obeys
 \begin{equation}
 \log |y_0| \mapsto \log |y_0| + \log C_2 + \delta_2\log |x_0|,
 \end{equation}
where $\delta_2$ is a function of $\beta$. The most positive normal Liapunov
exponent in this case is then
 \begin{equation}\label{eqlambdae}
 \Lambda = \log C_2 + \delta_2 \langle\log |x_0|\rangle,
 \end{equation}
where the average is taken over the Lorenz attractor. Averaging over $10^7$
iterates of~(\ref{eqlorenz}) yields $\langle\log |x_0|\rangle=-5.9233$ and,
solving (\ref{eqlambdae}) for $\beta$, we obtain $\beta=3.869$ at the blowout
bifurcation, in agreement with the data in Figure~\ref{figmapone}(b).

As $\Lambda(\mu_x)$ passes through $0$, $A_x$ loses stability and becomes a
chaotic saddle, and in doing so it creates a continuum of connections from
$A_x$ to the fixed point ($y$-rolls) in $R_y$. These connections are robust to
$G$-equivariant perturbations, as $y$-rolls are sinks within $Q_x$, and so
there is a robust cycle alternating between the equilibrium points in $R_x$ and
$R_y$ and the chaotic sets in $P_x$ and $P_y$.

We observed in sections~\ref{secnumerics} and \ref{secmap} that this cycling
chaos is attractive once it is created; we turn to the stability of cycling
chaos in the next section.

\section{Analysis of the resonance of cycling chaos}
\label{secresonance}

It is clear from Figures~\ref{figslowingdown}, \ref{figresonance} and
\ref{figmaptwo} that the key to understanding the loss of stability of cycling
chaos lies in obtaining the rate at which trajectories approach that state. It
is possible to estimate this rate for the map, and we use information gained in
this calculation to carry out the same estimate in the ODEs, and thus are able
to obtain the values of~$\beta$ at which the loss of stability occurs, in the
map and in the ODEs. What is remarkable is that, once a single average
over~$A_x$ has been computed numerically, the value of $\beta$ at the
bifurcation point can be obtained analytically.

\subsection{Rate of approach to cycling chaos}

We suppose that (in the map) the system arrives near~$P_x$ with given values of
$(x_0,1,y_0,y_2)$, iterates using~(\ref{eqnearPx}) $n$~times ($n\gg1$), ending
up in a state $(x_0',1,y_0',y_2')$ with $|y_0'|>|x_0'|$. There follows one
iterate of~(\ref{eqswitchPxPy}), leaving the system near~$P_y$ in a state
$(x_0'',x_2'',y_0'',1)$ (we are ignoring changes of sign of the variables). We
need to establish an estimate of $(x_0'',x_2'')$, a measure of the distance
from~$P_y$, given that $(y_0,y_2)$ were small when the system started close
to~$P_x$.

Properly, the value of $y_0'$ after $n$ iterates will depend on the values of
$x_0$ over those $n$ iterates, but if $n$ is large, we approximate the detailed
history of $x_0$ by its average and obtain
 \begin{eqnarray}
 \log|y_0'| &=& \log|y_0| + n\Lambda_{\mathrm{e}},       \\
 \log|y_2'| &=& \log|y_2| + n\Lambda_{\mathrm{c}},
 \end{eqnarray}
where
 \begin{eqnarray}
 \Lambda_{\mathrm{e}} &=& \log C_2 + \delta_2\langle\log|x_0|\rangle, \\
 \Lambda_{\mathrm{c}} &=& \log C_3 + \delta_3\langle\log|x_0|\rangle 
 \end{eqnarray}
are Liapunov exponents in the expanding and contracting directions
around~$P_x$. $\Lambda_{\mathrm{e}}$~is precisely the Liapunov exponent that
went through zero at the blowout bifurcation in~(\ref{eqlambdae}). Note that we
have effectively approximated the chaotic set $A_x$ by an equilibrium point.

The trajectory escapes from the neighbourhood of $P_x$ once $|y_0'|>|x_0'|$;
since $x_0$ is typically of order one (compared to the tiny initial value
of~$y_0$), we assume that the escape takes place when $|y_0'|=1$, so obtaining
 \begin{eqnarray}
 n &=& -\frac{\displaystyle 1}{\displaystyle\Lambda_{\mathrm{e}}}\log|y_0|, 
            \label{eqnumberits} \\
 \log|y_2'| &=& \log|y_2| - 
        \frac{\displaystyle\Lambda_{\mathrm{c}}}{\displaystyle\Lambda_{\mathrm{e}}}
        \log|y_0|,
 \end{eqnarray}
with $x_0'$ and $y_0'$ both of order one.

One iterate of~(\ref{eqswitchPxPy}) now yields
 \begin{eqnarray}
 \log|x_0''| &=& \log|x_0'| + \log C_4 
              + \delta_2 \log|y_0'|
              + \delta_4 \log|y_2'|,           \\
 \log|x_2''| &=&              \log C_5
              + \delta_3 \log|y_0'|
              + \delta_5 \log|y_2'|.
 \end{eqnarray}
These expressions will be dominated by $\log|y_2'|$ once the trajectory is very
close to the heteroclinic cycle, so, neglecting terms of order one, we obtain
 \begin{equation}\label{eqlinearisedmap}
 \left(\matrix{\log|x_0''|\cr
               \log|x_2''|\cr}\right)    =
 \left[\matrix{
 -\delta_4\frac{\displaystyle\Lambda_{\mathrm{c}}}{\displaystyle\Lambda_{\mathrm{e}}}&
  \delta_4\cr
 -\delta_5\frac{\displaystyle\Lambda_{\mathrm{c}}}{\displaystyle\Lambda_{\mathrm{e}}}&
  \delta_5\cr}\right]
 \left(\matrix{\log|y_0|\cr
               \log|y_2|\cr}\right) + {\cal O}(1)
 \end{equation}
for large negative values of $\log|y_0|$ and $\log|y_2|$. A conjugate map will
describe the return from $P_y$ to $P_x$. One eigenvalue of the matrix is zero
because of the way we approximated the behaviour near the chaotic set; the
other eigenvalue is
 \begin{equation}\label{eqrhomap}
 \rho = \delta_5 
      - \delta_4 \frac{\displaystyle\Lambda_{\mathrm{c}}}
                      {\displaystyle\Lambda_{\mathrm{e}}},
 \end{equation}
which we refer to as the {\em stability index}. 

The zero eigenvalue forces $\log|x_0''|=(\delta_4/\delta_5)\log|x_2''|$, so
after one iterate of the composite map~(\ref{eqlinearisedmap}), the dynamics
will obey
 \begin{equation}\label{eqlinearisedmapprojected}
 \left(\matrix{\log|x_0''|\cr
               \log|x_2''|\cr}\right)    =
 \rho
 \left(\matrix{\log|y_0|\cr
               \log|y_2|\cr}\right).
 \end{equation}
Clearly if $\rho>1$, $\log|x_0|$ and $\log|y_0|$ will tend to $-\infty$ and the
trajectory will asymptote to attracting cycling chaos. Conversely, if $\rho<1$,
cycling chaos is unstable and trajectories will leave the domain of validity of
the approximations we have made. We can also deduce from~(\ref{eqnumberits})
that the number of iterates between each switch will increase by a factor
of~$\rho$ per switch.

At the point at which cycling chaos is created (as $\Lambda_{\mathrm{e}}$
increases through zero), we see that $\rho$ is much greater than one, provided
that $\Lambda_{\mathrm{c}}$ is negative and $\delta_4$ is positive, both of
which are true in the examples we have discussed. We deduce that cycling chaos
is attracting near to its creation at the blowout bifurcation.

\subsection{Loss of stability of cycling chaos}

 From the condition that $\rho=1$ at the loss of stability of the chaotic
cycle, we determine that this bifurcation occurs in the map at $\beta=1.0896$,
in agreement with the data in Figure~\ref{figmaptwo}.

Returning to the ODEs, we observe that in (\ref{eqrhomap}) $\delta_4$ and
$\delta_5$ are ratios of eigenvalues of $x$-rolls (proportional to decay rates
of $y_0$ and~$y_2$), while $\Lambda_{\mathrm{e}}$ and $\Lambda_{\mathrm{c}}$
are the growth rate of $y_0$ and the decay rate of $y_2$ near the chaotic
set~$A_x$. In the ODEs, the linearisation of~(\ref{eqode}) about~$A_x$ yields
$\lambda_\zeta$ for the decay rate of~$y_2$, while the growth rate of~$y_0$ is
given by~(\ref{eqlambdae}). Hence we have the stability index
 \begin{equation}\label{eqrhoodes}
 \rho = \delta_5 
      - \delta_4 \frac{\displaystyle\lambda_\zeta}
                      {\displaystyle\Lambda_{\mathrm{e}}}
 \end{equation}
 for the ODEs, where $\Lambda_{\mathrm{e}}$ is given by~(\ref{eqlambdae}). Note
that $\delta_4$ and $\Lambda_{\mathrm{e}}$ are both functions of~$\beta$. The
condition $\rho=1$ is readily solved for $\beta$, and has solution
$\beta=1.63$. At $\beta=1.63$ the ODEs are still approaching cycling chaos,
with switching times increasing by a factor of above 1.02 per switch (see
Figure~\ref{figresonance}). However, the ODEs have not yet reached their
asymptotic rate of slowing down, resolving this small discrepancy.

On decreasing $\beta$ below $1.63$, the stability index~$\rho$ increases above
unity and the cycling chaos will no longer be attracting. This loss of
stability can broadly be classed as a resonance of Liapunov exponents. Observe
that the resonance will be located at different $\beta$ for different invariant
measures supported on $A_{x,y}$ and so we expect the presence of riddling and
associated phenomena found in \cite{Ash97} at a resonance of a simpler model
displaying cycling chaos.

We observe that for $\beta$ below the resonance bifurcation, the system
exhibits behaviour that is numerically indistinguishable from periodic: there
appear to be a large number of coexisting apparently periodic orbits. We
hypothesize that the resonance creates a branch of `approximately periodic
attractors', i.e., attractors that have a well-defined finite mean period of
passage around the cycle, going to infinity at the resonance \cite{Ash97}.
These might lock onto long periodic orbits for progressively smaller $\beta$,
as found in the numerical simulations. For this example, the approximately
periodic attractor branches set supercritically from the cycling chaos; however
one presumes that subcritical branching is also possible. Research is presently
in progress on understanding the more detailed branching behaviour at this
bifurcation.

\section{Discussion}
\label{secdiscuss}

This study has shown that one possible, apparently generic scenario for loss of
stability of a chaotic attractor in an invariant subspace on varying a normal
parameter is as follows: there is a blowout bifurcation that creates an
attracting, robust heteroclinic cycle between chaotic invariant sets (cycling
chaos). The bifurcation is set supercritical but not supercritical, i.e., the
bifurcated attractors contain the attractor for the system in the invariant
subspace, but unlike in a supercritical bifurcation (to an on-off intermittent
state) the length of laminar phases increases unboundedly along a single
trajectory even at a finite distance from the blowout.

This cycling chaotic state can be modelled well by the network shown in
Figure~\ref{figcycle} although in reality the network is complicated by the
facts that (a) there are other fixed points contained in the closure of
unstable manifolds and (b) the fixed points in $F_x$ and $F_y$ are actually
contained in the chaotic sets $A_x$ and $A_y$ rather than being isolated. We
suspect this may have the consequence that there is no Poincar\'{e} section to
the flow and so the cycle is `dirty' in the terminology of \cite{Ash97}.
Nevertheless, the normal Liapunov spectrum of the invariant chaotic set seems
to determine the attraction or not of the cycle.

The attracting cycling chaos is observed to lose stability via a mechanism that
resembles a resonance of eigenvalues in an orbit heteroclinic to equilibria.
Such a resonance has been seen to occur in special classes of systems with skew
product structure \cite{Ash97}, in analogy to the branching of periodic orbits
at a homoclinic resonance investigated by \cite{Cho&al90}.

Throughout this investigation, it has been necessary to monitor carefully
several numerical effects. In particular, for trajectories that display
asymptotic slowing down characteristic of cycling chaos there will be a point
at which rounding errors cause the dynamics either to transfer to the invariant
subspace, or keep the dynamics a finite distance from the invariant subspace.
In the context of physical systems there will always be imperfections in the
system and noise that will destroy the invariant subspaces. Nevertheless the
perfect symmetry model will be very useful in describing what one expects to
see in such imperfect situations.

It still remains to prove rigorously that the observed scenario is generic and
so of interest to other, less specific models and in particular PDE models of
which this is a truncation.  We could like to emphasise that the behaviour we
see occurs for a reasonably large region of physically relevant parameters in
the ODE model and moreover we are unaware of any other classification which
explains and predicts the observed dynamics to the degree that we have done
here. In principal cycling chaos can be seen in ODE models down to dimension 4,
though not smaller than this; thus this dynamics should be seen as something
that will not be created at a generic bifurcation from a trivial state, but
rather in a more complicated dynamical regime far from primary bifurcation.

We have discussed a possible route to cycling chaos through a blowout
bifurcation, and how cycling chaos might cease to be attracting, both in
general terms and in the context of a specific model. Our general results ought
to be applicable to a variety of other problems. In particular,
behaviour that might be understood in terms of cycling chaos has been seen by
Knobloch {\em et al.}\ \cite{KTW} in an ODE model of the dipole--quadrupole
interaction in the solar dynamo. In their model, there is a weakly broken
symmetry between the dipole and quadrupole subspaces, and the system switches
between the two subspaces, favouring one over the other since they are not
equivalent.

Finally, one comment that deserves to be made is that the choice of $\beta$ as
the parameter allows an important simplification because of this parameter is
normal for dynamics within $P_x$ and $P_y$. One assumes that similar behaviour
will be observed for non-normal parameters with the exception that the chaos in
the invariant subspace will be fragile \cite{Bar&al97} and destroyed by many
arbitrarily small perturbations; see \cite{Cov&al97}.

 \begin{ack}
 We acknowledge very interesting conversations with Mike Field, Marty
Golubitsky and Edgar Knobloch concerning this work. The research of PA was
partially supported by a Nuffield `Newly Appointed Science Lecturer' award and
EPSRC grant GR/K77365. AMR is grateful for support from the Royal Astronomical
Society.
 \end{ack}


\begin{thebibliography}{99}

\bibitem{Ale&al92} 
J.C.~Alexander, I.~Kan, J.A.~Yorke and Zhiping~You.
\newblock Riddled Basins.
\newblock {\em Int.\ Journal of Bifurcations and Chaos} 
          {\bf 2} (1992) 795--813.

\bibitem{Ash97}
P.~Ashwin.
\newblock Cycles homoclinic to chaotic sets; robustness and resonance.
\newblock {\em Chaos} 
          {\bf 7} (1997) 207--220.

\bibitem{AshAstNic96}
P.~Ashwin, P.~Aston and M.~Nicol.
\newblock On the unfolding of a blowout bifurcation.
\newblock {\em Physica D} to appear (1997).

\bibitem{AshBueSte94}
P.~Ashwin, J.~Buescu and I.N.~Stewart.
\newblock Bubbling of attractors and synchronisation of oscillators.
\newblock {\em Phys.\ Lett.~A} 
          {\bf 193} (1994) 126--139.

\bibitem{AshBueSte96}
P.~Ashwin, J.~Buescu and I.N.~Stewart.
\newblock From attractor to chaotic saddle: a tale of transverse instability.
\newblock {\em Nonlinearity} 
          {\bf 9} (1996) 703--737.

\bibitem{Bar&al97}
E.~Barreto, B.~Hunt, C.~Grebogi and J.~Yorke.
\newblock From high dimensional chaos to stable periodic orbits: the structure of parameter space.
\newblock {\em Phys.\ Rev.\ Lett.} to appear (1997).

\bibitem{Cho&al90}
S.-N.~Chow, B.~Deng and B.~Fiedler.
\newblock Homoclinic bifurcation at resonant eigenvalues.
\newblock {\em J.\ Dyn.\ Diff.\ Eqns.} 
          {\bf 2} (1990) 177--244.

\bibitem{Cov&al97}
E.~Covas, P.~Ashwin and R.~Tavakol.
\newblock {\em Non-normal parameter blowout bifurcation in a truncated dynamo model}
\newblock Preprint, University of Surrey (1997).

\bibitem{Del&al95}
M.~Dellnitz, M.~Field, M.~Golubitsky, A.~Hohmann and J.~Ma.
\newblock Cycling chaos.
\newblock {\em IEEE Trans.\ Circuits and Systems-I} 
          {\bf 42} (1995) 821--823.

\bibitem{Fie96}
M.~Field.
\newblock {\em Lectures on bifurcations, dynamics and symmetry}.
\newblock Pitman Research Notes in Mathematics, {\bf 356}, 1996, Pitman.

\bibitem{Guc&Hol88}
J.~Guckenheimer and P.~Holmes.
\newblock Structurally stable heteroclinic cycles.
\newblock {\em Math.\ Proc.\ Camb.\ Phil.\ Soc.} 
          {\bf 103} (1988) 189--192.
 
\bibitem{KTW}
E.~Knobloch, S.M.~Tobias and N.O.~Weiss.
\newblock Modulation and symmetry changes in stellar dynamos.
\newblock {\em Mon.\ Not.\ Roy.\ Astr.\ Soc.} to be submitted (1997).

\bibitem{KrupaMelbourne}
M.~Krupa and I.~Melbourne.
\newblock Asymptotic stability of heteroclinic cycles in systems with symmetry
\newblock {\em Ergod.\ Th.\ \& Dynam.\ Sys.} 
          {\bf 15} (1995) 121--147.
 
\bibitem{Metal96}
P.C.~Matthews, A.M.~Rucklidge, N.O.~Weiss and M.R.E.~Proctor.
\newblock The three-dimensional development of the shearing instability of convection.
\newblock {\em Phys.\ Fluids} 
          {\bf 8} (1996) 1350--1352.

\bibitem{Mil85}
J.~Milnor.
\newblock {On the concept of attractor.}
\newblock {\em Commun.\ Math.\ Phys.} {\bf 99} (1985) 177--195;
\newblock Comments {\em Commun.\ Math.\ Phys.} {\bf 102} (1985) 517--519.

\bibitem{Ott&Som94}
E.~Ott and J.C.~Sommerer.
\newblock Blowout bifurcations: the occurrence of riddled basins and on-off
intermittency.
\newblock {\em Phys.\ Lett.~A} 
          {\bf 188} (1994) 39--47.

\bibitem{Pla&al93}
N.~Platt, E.A.~Spiegel and C.~Tresser.
\newblock On-off intermittency; a mechanism for bursting.
\newblock {\em Phys.\ Rev.\ Lett.} 
          {\bf 70} (1993) 279--282.

\bibitem{NumRecipes}
W.H.~Press, B.P.~Flannery, S.A.~Teukolsky and W.T.~Vetterling
\newblock {\em Numerical Recipes -- the Art of Scientific Computing}.
\newblock (Cambridge University Press, Cambridge, 1986).
 
\bibitem{RM95}
A.M.~Rucklidge and P.C.~Matthews.
\newblock The shearing instability in magnetoconvection, in
\newblock {\em Double-Diffusive Convection}, 
          eds.\ A.~Brandt and H.J.S.~Fernando) 
          (American Geophysical Union, Washington, 1995) pp.~171--184.

\bibitem{RM96}
A.M.~Rucklidge and P.C.~Matthews.
\newblock Analysis of the shearing instability in nonlinear convection and magnetoconvection
\newblock {\em Nonlinearity} 
          {\bf 9} (1996) 311--351.

\bibitem{Sig92}
K.~Sigmund.
\newblock Time averages for unpredictable orbits of deterministic systems.
\newblock {\em Annals of Operations Research} 
          {\bf 37} (1992) 217--228.
 
\end{thebibliography}
\end{document}